\documentclass[floatfix,pre,showpacs,preprint,superscriptaddress]{revtex4-1}

\usepackage{graphicx}
\usepackage{epstopdf}

\usepackage{amsmath}
\usepackage{amssymb}

\usepackage{bm}

\usepackage{latexsym}

\usepackage{verbatim}
\usepackage[usenames]{color}
\usepackage{xspace}

\def\eg{\textit{e.g.}\xspace}

\def\kt{k_\text{B}T}
\def\Lt{L_\text{tran}}
\def\lamb{\lambda_\text{b}}
\def\lamr{\lambda_\text{r}}
\def\Cb{C_\text{b}}
\def\Cr{C_\text{r}}
\def\qb{q_\text{b}}
\def\qr{q_\text{r}}

\begin{document}

\title{Interaction of chiral rafts in self-assembled colloidal membranes }

\author{Sheng Xie}
\affiliation{Department of Physics, Brown University, Providence RI, 02912, U.S.A}
\author{Michael F. Hagan}\email{hagan@brandeis.edu}
\affiliation{Department of Physics, Brandeis University, Waltham MA, 02454, U.S.A.}
 \author{Robert A. Pelcovits}\email{pelcovits@brown.edu}
 \affiliation{Department of Physics, Brown University, Providence RI, 02912, U.S.A}

\date{\today}

\begin{abstract}
Colloidal membranes are monolayer assemblies of rodlike particles that capture the long-wavelength properties of lipid bilayer membranes on the colloidal scale. Recent experiments on colloidal membranes formed by chiral rodlike viruses showed that introducing a second species of virus with different length and opposite chirality leads to the formation of rafts --- micron-sized domains of one virus species floating in a background of the other viruses [Sharma et al., Nature 513, 77 (2014)].
In this article we study the interaction of such rafts using liquid crystal elasticity theory. By numerically minimizing the director elastic free energy, we predict the tilt angle profile for both a single raft and two rafts in a background membrane, and the interaction between two rafts as a function of their separation. We find that the chiral penetration depth in the background membrane sets the scale for the range of the interaction.  We compare our results with the experimental data and find good agreement for the strength and range of the interaction. Unlike the experiments, however, we do not observe a complete collapse of the data when rescaled by the tilt angle at the raft edge.
\end{abstract}

\maketitle

\section{INTRODUCTION}
\label{intro}

Placing a system of monodisperse rodlike viruses (\eg filamentous bacteriophages) in the presence of a non-adsorbing polymer leads to effective depletion attractions that drive lateral association of rods. When the rodlike particles are chiral, their lateral interactions have a preferred twist, causing them to form a variety of fascinating one-rod-length thick assemblies, including one-dimensional twisted ribbons, braided ribbons, starfish shapes and membranes with twist domain walls \cite{Dogic2014, Barry2010, Gibaud2012,Yang2012,Yang2011, Barry2009a, Barry2009, Zakhary2014, Kang2015,Dogic2014,Kaplan2014,Kaplan2013}.  Considerable theoretical effort has been devoted to modeling and understanding the properties of these self-assemblies (e.g., \cite{Yang2012, Yang2011, Barry2009, Kang2015, Kaplan2014, Kaplan2013, Savenko2006,Patti2009,Cuetos2010,Cuetos2008,Pelcovits2009,Kang2015, Kaplan2010,Tu2013,Tu2013a}).

Mixing rods of two different lengths yields yet more interesting structures. Sharma et al. \cite{key-41} recently studied mixtures of the filamentous bacteriophages \textit{fd}-Y21M and M13KO7.  The two viruses have the same diameter, but \textit{fd}-Y21M virus has a contour length of 880 nm and has right-handed chirality, while M13KO7 has a contour length of 1,200 nm and exhibits left-handed chirality \cite{Barry2009a, Tombolato}. Over a broad range of polymer depletant concentrations, the two rod species co-assemble to form bidisperse membranes (still one-rod-length thick), but the distributions of the rods within the membrane depend sensitively on polymer concentration. At low polymer concentrations rods mix homogeneously, while at high concentrations they undergo macrophase separation. At intermediate polymer concentrations, Sharma et al. observed equilibrium microphase separation of the two viruses into highly monodisperse micron-sized colloidal rafts composed of the shorter \textit{fd}-Y21M virus, floating in a background membrane composed primarily of the longer M13KO7 viruses.

By manipulating the positions and sizes of rafts with optical tweezers, Sharma et al. \cite{key-41} measured the interaction potential between pairs of rafts, which they found to be repulsive and exponentially decaying with raft edge-edge separation.
They suggested a physical mechanism for this repulsion, noting that the formation of right-handed rafts in a background of left-handed rods helps alleviate the frustration that arises from the incompatibility of chiral twist and membrane layer formation. They provided evidence for this explanation by measuring the rod tilt angle at the edge of the rafts, and found that the raft interaction potential for rafts of different sizes collapses on top of each other when rescaled by the tilt angle at the edge. Furthermore, they found that the range of the interaction is independent of raft size, while the strength grows linearly with increasing raft size.

In this paper we present quantitative modeling of the physical mechanism suggested by Sharma et al., focusing on the role of liquid crystal director elasticity in the interaction of chiral rafts. We do not address the formation or optimal size of the rafts. We assume that rafts of a particular size have formed (either naturally or via manipulation by optical tweezers), and determine the tilt angle in their interiors and the intervening background membrane, by numerically minimizing the Frank elastic energy. We then compute the interaction between the rafts by comparing the free energy of the system (rafts plus background membrane) with a reference system in which the rafts are far apart and noninteracting. Our tilt angle profiles support the physical mechanism for the repulsive raft interaction suggested by Sharma et al. \cite{key-41}. The theory allows prediction of how the strength of the raft interaction potential depends on a number of experimentally controllable parameters, such as raft size, the magnitude of chiral twist and elastic moduli for raft or background rods. We find that the twist penetration depth in the background membrane sets the characteristic decay length of the the raft interaction. Thus, as observed in the experiments, we find that the decay length is independent of raft diameter. Also in agreement with experiment, we find that the strength of the raft interaction increases with increasing raft diameter.  One discrepancy between theory and  experiments is that experimentally measured interaction potentials for different-sized rafts collapse into a single curve when  rescaled by the tilt angle at the raft edge \cite{key-41}; our theoretical curves come closer together upon rescaling but do not fully collapse.

The mediation of the raft interaction by the background director field is an example of a more general phenomenon of membrane-mediated interactions between membrane inclusions, of which there are examples in liquid crystal systems \cite{Poulin, Pettey, Dolganov, Bohley}, models of embedded proteins in biological  membranes \cite{Dan, Goulian} and lipid rafts in cell membranes \cite{Dietrich, Lingwood}. The bidisperse filamentous bacteriophage experiments established a highly controllable experimental platform with which to measure such membrane-mediated interactions. Our theory provides a means to probe the physical mechanisms underlying these measurements.

This paper is organized as follows. In the next section we describe the director free energy model used in our analysis, and our numerical method for its minimization.  In  Sec.~\ref{results} we first present the predicted tilt angle profile of a single raft in a background membrane of opposite chirality, followed by the tilt angle profiles and interaction potential for two such rafts. We compare our results with the experimental data of Ref.~\cite{key-41}.  Sec.~\ref{conclusions} offers concluding remarks. The appendices contain technical details of the discretization of the free energy and a calculation of the tilt angle at the interface between two semi-infinite membranes of different chirality.

\section{Free energy and its minimization}

We consider a single, flat smectic layer of chiral $fd$ viruses whose long axes are specified by a director field $\hat{\mathbf n}$. The de Gennes free energy density for this system in the single Frank constant approximation is given by \cite{deGennesSmectic,PelcovitsMeyer}:
\begin{equation}
f_\text{n}=\frac{1}{2}K[(\nabla\cdot\hat{\mathbf n})^{2}-2q\hat{\mathbf n}\cdot(\nabla\times\hat{\mathbf n})+(\nabla\times\hat{\mathbf n})^{2}+q^{2}]+\frac{1}{2}C\sin^{2}\theta\label{eq:4.3}.
\end{equation}
where $K$ is the Frank elastic constant, $q$ is the chirality, $\theta$ is the tilt angle of the director
with respect to the membrane normal and $C$ is the tilt energy modulus.
Director twist is expelled from the interior of the membrane, but can penetrate near the edges over a distance $\lambda = \sqrt{K/C}$, the twist penetration depth \cite{deGennesSmectic}. To model virus mixtures we will allow $q$ and $C$ to depend on position, taking on values $\qr$ and $\Cr$ in the rafts and $\qb$ and $\Cb$ in the background membrane.
The three dimensional molecular director field $\hat{\mathbf n}$ can be expressed in terms of $\theta$ and $\varphi$ (the azimuthal angle in the plane of the membrane) by:
\begin{equation}
\hat{\mathbf n}=(n_{x},n_{y},n_{z})=(\sin\theta \cos\varphi,\sin\theta \sin\varphi,\cos\theta)\label{eq:4.4}.
\end{equation}

With the membrane normal chosen along the \textit{z} axis,  $f_{n}$ is given by:
\begin{eqnarray}
f_\text{n}=\frac{K}{2}&\Biggl\lbrack&\biggl(\frac{\partial\theta}{\partial x}\biggr)^2+\biggl(\frac{\partial\theta}{\partial y}\biggr)^{2}+\sin^{2}\theta\biggl(\biggl(\frac{\partial\varphi}{\partial x}\biggr)^{2}+\biggl(\frac{\partial\varphi}{\partial y}\biggr)^2\biggr)+2\cos\theta \sin\theta\biggl(\frac{\partial\theta}{\partial x}\frac{\partial\varphi}{\partial y}-\frac{\partial\theta}{\partial y}\frac{\partial\varphi}{\partial x}\biggr)+q^{2}\Biggr\rbrack\nonumber\\
&&{}-Kq\biggl\lbrack \sin\varphi\frac{\partial\theta}{\partial x}-\cos\varphi\frac{\partial\theta}{\partial y}+\sin\theta \cos\theta\biggl(\cos\varphi\frac{\partial\varphi}{\partial x}+\sin\varphi\frac{\partial\varphi}{\partial y}\biggr)\biggr\rbrack+\frac{C}{2}\sin^{2}\theta\label{fn}
\end{eqnarray}

We discretize the free energy density (see Appendix A for details) and then use the Newton-Gauss-Seidel method \citep{key-76} to minimize its value. The Newton-Gauss-Seidel method is specified by \cite{Stark}:
\begin{align}
\theta^\text{new}(x,y)&=\theta^\text{old}(x,y)-\frac{\delta f_\text{n}/\delta\theta(x,y)}{`` \delta^2 f_\text{n}/\delta^2 \theta(x,y)"} \\
\varphi^\text{new}(x,y)&=\varphi^\text{old}(x,y)-\frac{\delta f_\text{n}/\delta\varphi(x,y)}{`` \delta^2 f_\text{n}/\delta^2 \varphi(x,y)" }\label{eq:4.8b}
 \end{align}
where ``��$\delta^{2}f_\text{n}/\delta^{2}\theta(x,y)$'' is calculated as the derivative of $\delta f_\text{n}/\delta\theta(x,y)$
with respect to $\theta(x,y)$ at a grid point $(x,y)$ and similarly for $\varphi$. The quotation marks indicate that these derivatives are not the discretized form of a real second-order functional derivative, which would involve a delta function \cite{Stark}. The iteration is repeated until the change in the total free energy of the system is less than one part in $10^7$.

\begin{figure}
\includegraphics[scale=0.5]{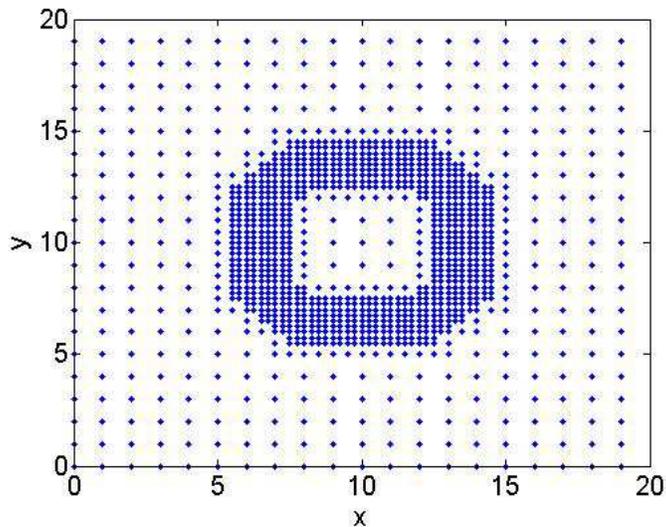}
\protect\caption{Illustration of mesh refinement for a single raft of radius 4 centered at (10,10). A much finer grid was utilized in obtaining our results. \label{fig:A-simple-grid}}
\end{figure}

We consider a grid of points in a square box of side $L=60$ (measured in units of the twist penetration
depth $\lambda$) with periodic boundary conditions. The box is meant to represent a small portion of the membrane far from its edge. The initial spacing between neighboring grid points is 0.5, yielding a grid of $120\times120$ points. Within this box we place two rafts, defined as discrete circular areas of radius $R$. For grid points lying inside the rafts $q=\qr$, while for points outside $q= \qb$, where $\qr$ and $\qb$ have opposite signs and possibly different magnitudes. We initialize the director at each grid point in the box as a small random fluctuation about the $z$ axis and relax the system using Eq.~(\ref{eq:4.8b}). After a few thousand steps (each step being a loop over all grid points), the grid close to the edges
of the rafts is refined by adaptive mesh refinement \citep{key-67}. Figure \ref{fig:A-simple-grid} shows a simple example of mesh refinement applied to a single raft. After the system equilibrates, the interaction potential (measured in units of $K$) between the two rafts is extracted by comparing
the total free energy of the system with that of a reference system where the two rafts are far enough apart that no significant interaction exists.

\section{Results}
\label{results}

\subsection{Single raft\label{sub:Single-raft}}

\begin{figure}
\includegraphics[scale=0.5]{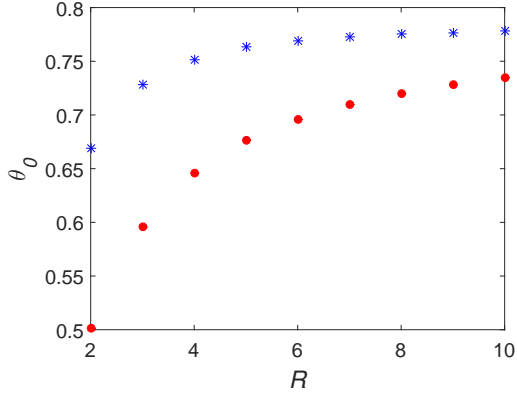}
\protect\caption{Tilt angle $\theta_{0}$ (stars) at the edge of a single raft as a function of
the raft radius $R$ (measured in units of  $\lambda$) with $q=0.71$ in a background membrane with chirality equal in magnitude and opposite in sign. The solid circles are the corresponding tilt angle values from Ref. \cite{PelcovitsMeyer} for circular membranes surrounded by depletant. \label{fig:a)-Plot-of}}
\end{figure}

We first investigate a single raft of radius $R$ with chirality equal in magnitude and opposite in sign to the surrounding background, i.e., $\qb=-\qr=q$. We assume that the tilt modulus $C$ has the same value within the raft and the background membrane, i.e., $\Cr=\Cb$; thus, the penetration depth $\lambda=\sqrt{K/C}$ has a common value as well.  For simplicity we assume that the sign of the chirality changes abruptly at the edge of the raft; we consider other chirality  profiles when we study two rafts. Figure \ref{fig:a)-Plot-of} shows the tilt angle $\theta_{0}$ at the raft boundary as a function of $R$ (measured in units of $\lambda$). We have chosen $q=0.71$, the value used in Ref.~\cite{PelcovitsMeyer}, where circular membranes bounded by the polymer depletant solution were studied by numerically solving the Euler-Lagrange equation for $f_\text{n}$. In that case the relevant boundary conditions were zero tilt at the center of the membrane and zero director torque at the edge. From the figure we see that, in the case of the raft surrounded by a chiral membrane of opposite chirality, the value of $\theta_0$ is larger than in similarly sized circular membranes surrounded by depletant \cite{PelcovitsMeyer}, although this difference decreases as $R$ increases. This decrease at large $R$ is consistent with an analytic calculation (see Appendix B) of $\theta_0$ at the interface between two \textit{semi-infinite} membranes of equal and opposite chirality. This calculation yields the same value for $\theta_0$ as found in Ref.~\cite{Barry2009} for a semi-infinite membrane bounded by a polymer depletant with zero director torque at the boundary.

\begin{figure}
\includegraphics[scale=0.4]{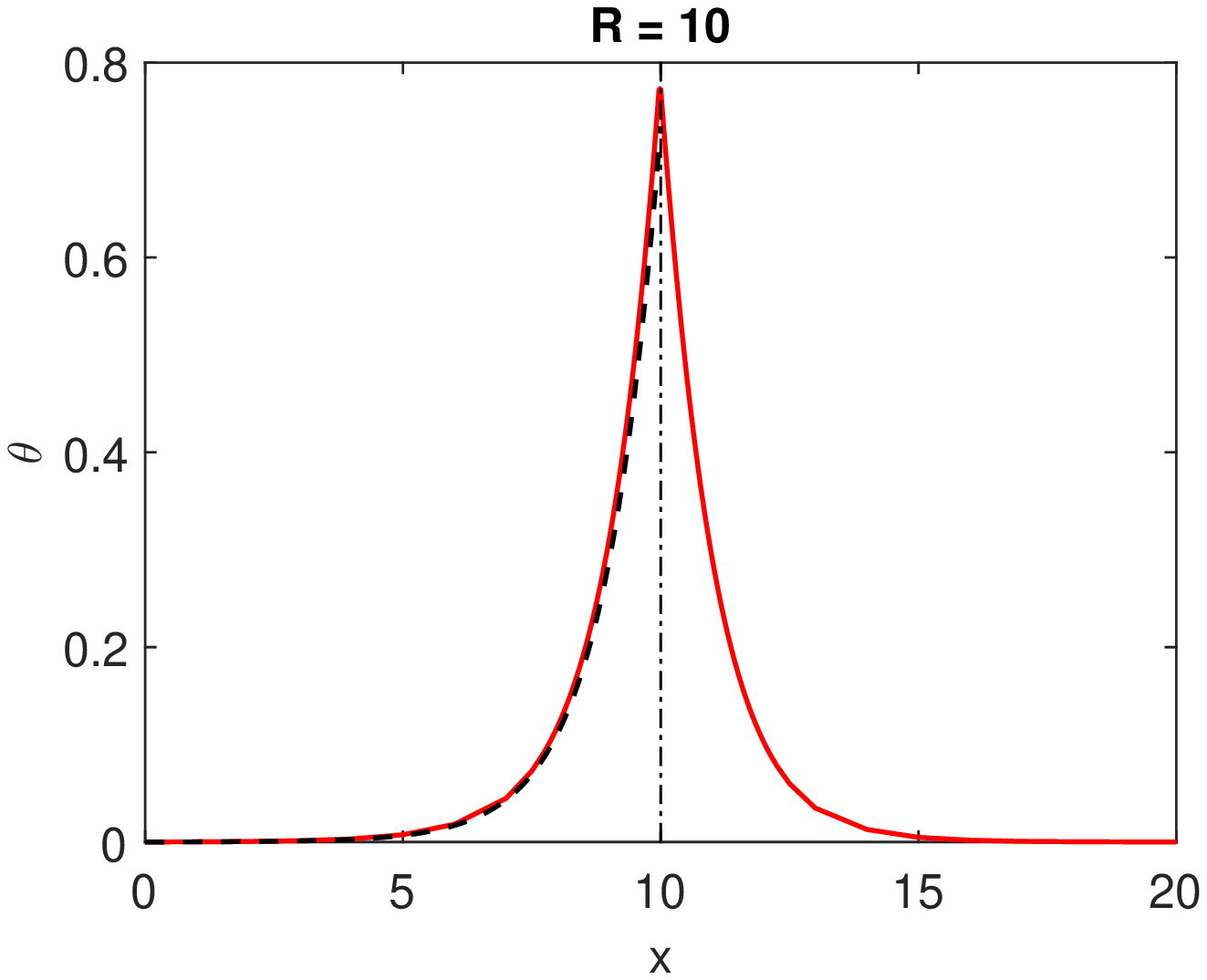}
\includegraphics[scale=0.4]{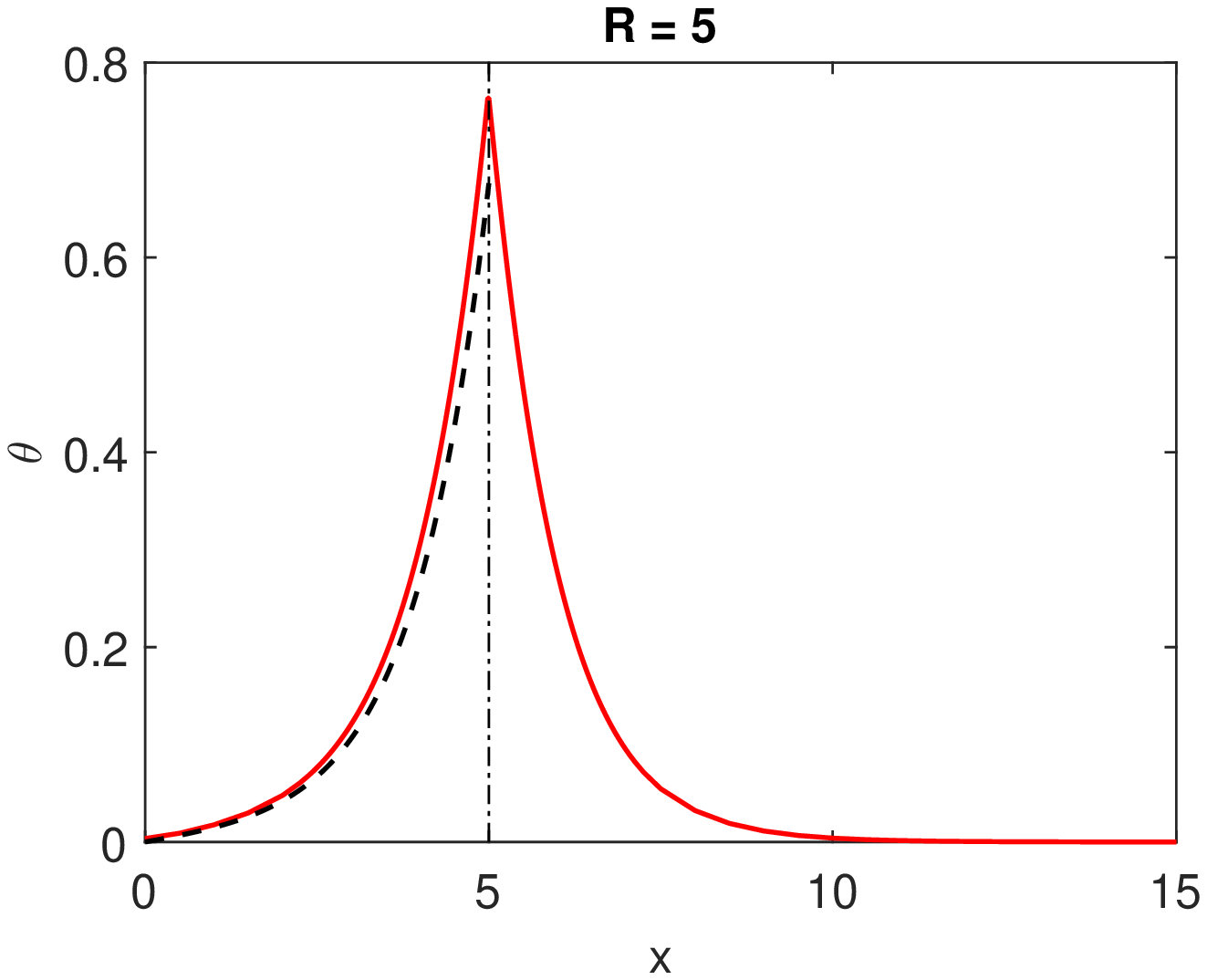}
\includegraphics[scale=0.4]{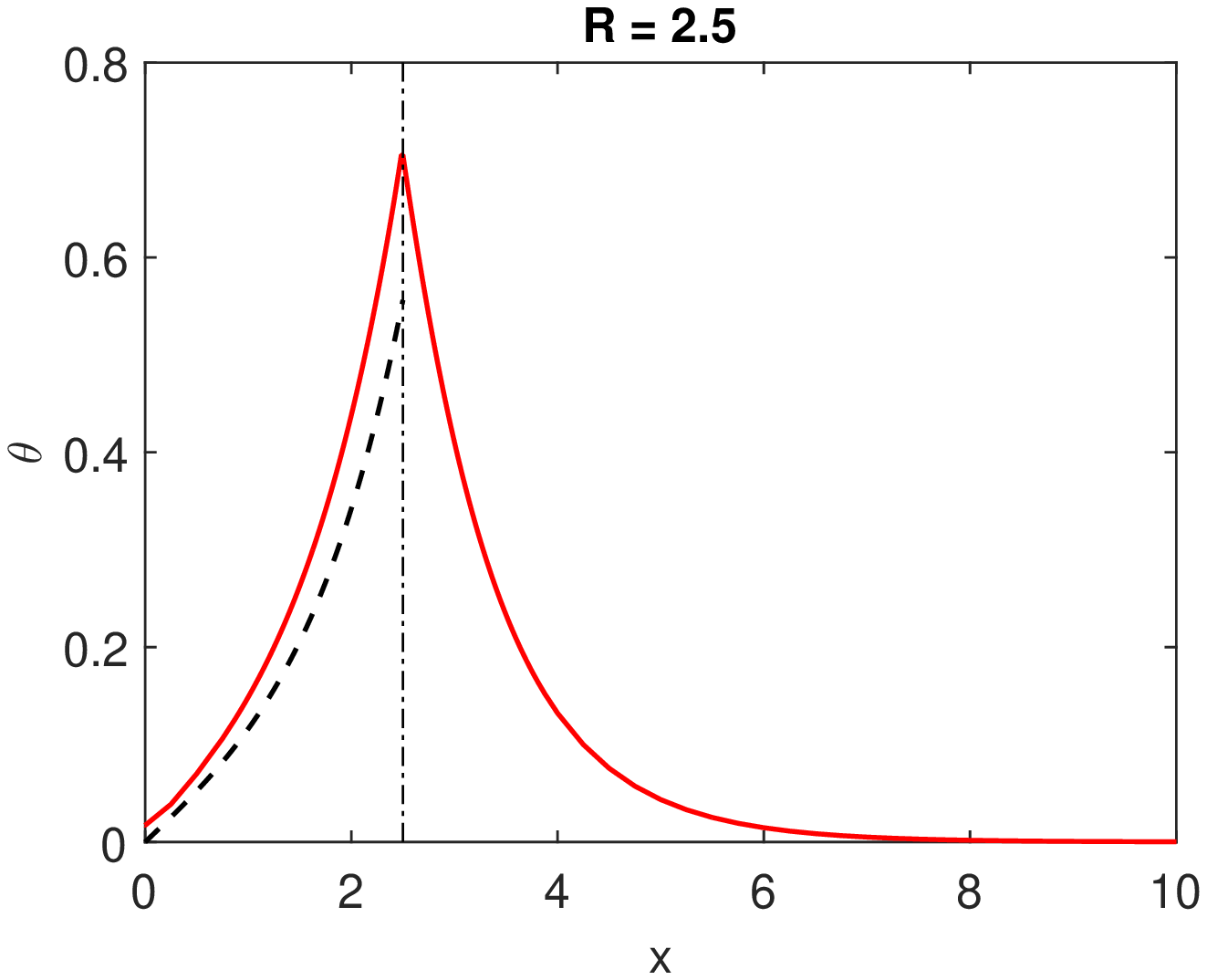}
\includegraphics[scale=0.4]{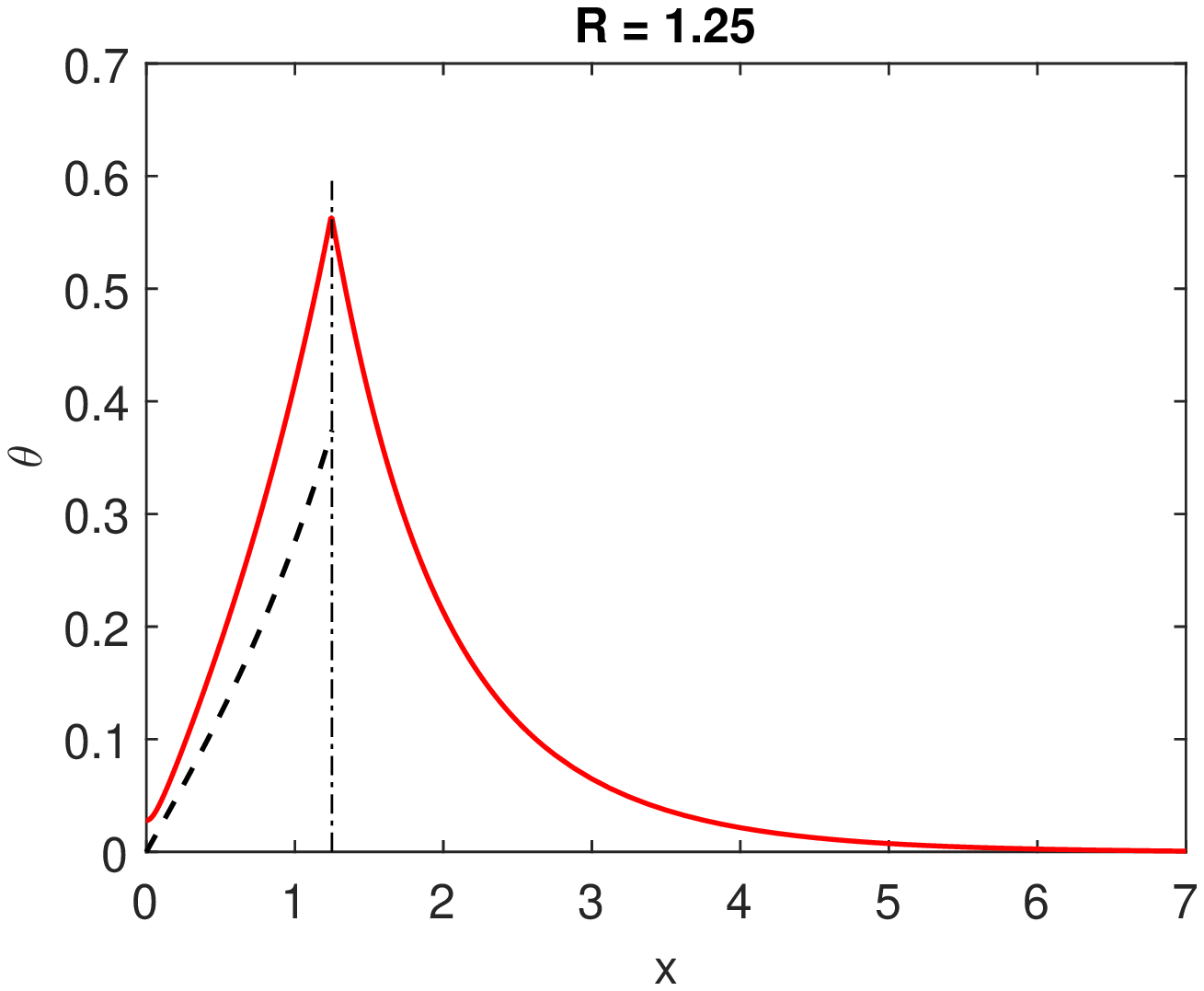}
\protect\caption{ Tilt angle $\theta$ (solid line) in a single raft surrounded by a background membrane of equal and opposite chirality ($q=0.71$) as a function of the $x$ coordinate (units of $\lambda$) measured from the center of the raft. The values of $\theta$ are compared with the results (dashed curve) found in Ref.~\cite{PelcovitsMeyer} for circular membranes surrounded by depletant.  The centers of the raft and the circular membrane are both located at the origin. The vertical dashed-dotted line denotes the edge of the raft and circular membrane. \label{fig:Plots-of-tilt}}
\end{figure}

Figure \ref{fig:Plots-of-tilt} shows the tilt angle $\theta$
through a cross section of the raft and background membrane.  In the figure we compare our results to those obtained in Ref.~\cite{PelcovitsMeyer}  for  circular membranes surrounded by depletant.  From the figure we see that as  $R$ increases, the $\theta$ profile for the raft becomes more symmetric about the edge of the raft and is almost perfectly symmetric once $R\gtrsim 5$ and  $\theta$ decays nearly to zero at the center of the raft. This approximate critical value of $R$ was also noted in Ref.~\cite{PelcovitsMeyer} for the tilt profile in isolated circular membranes. From the figure we also see how the system accommodates the opposite chiralities of the raft and background.  Starting from the center of the raft and moving outward, the tilt angle grows, following the twist preferred by the raft rods, until reaching the edge of the raft. As one moves into the background membrane where the viruses have opposite chirality, the rods twist in the opposite sense and the tilt angle decreases.

\begin{figure}
\includegraphics[scale=0.45]{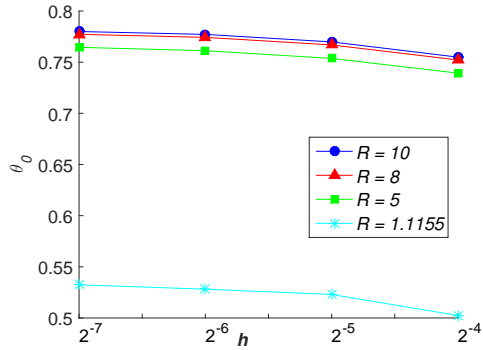}
\protect\caption{ Dependence of  $\theta_{0}$, the tilt angle at the raft edge, on mesh
refinement $h$ with $q=0.71$ for four different values of raft radius (measured in units of $\lambda$). The horizontal axis is the finest mesh
size level (i.e., near the raft edge), measured in units of $\lambda$.   \label{fig:raft mesh sensitivity}}
\end{figure}

To check the accuracy of our model, we computed $\theta_0$, the tilt angle at the edge of the raft, for different mesh sizes. Our results are shown in Figure
\ref{fig:raft mesh sensitivity}, where the horizontal axis shows the smallest mesh size at each
refinement level, and the value of $\theta_{0}$ is taken as an indicator of the influence of refinement levels. The
initial coarse mesh size is 0.5. Three different raft sizes were tested,
and the result shows that $\theta_{0}$ converges at the same rate
for all three raft sizes as the mesh size is decreased, which indicates that
our method is numerically consistent.

\subsection{Two rafts\label{sub:Two-rafts}}

\begin{figure}
\includegraphics[scale=0.5]{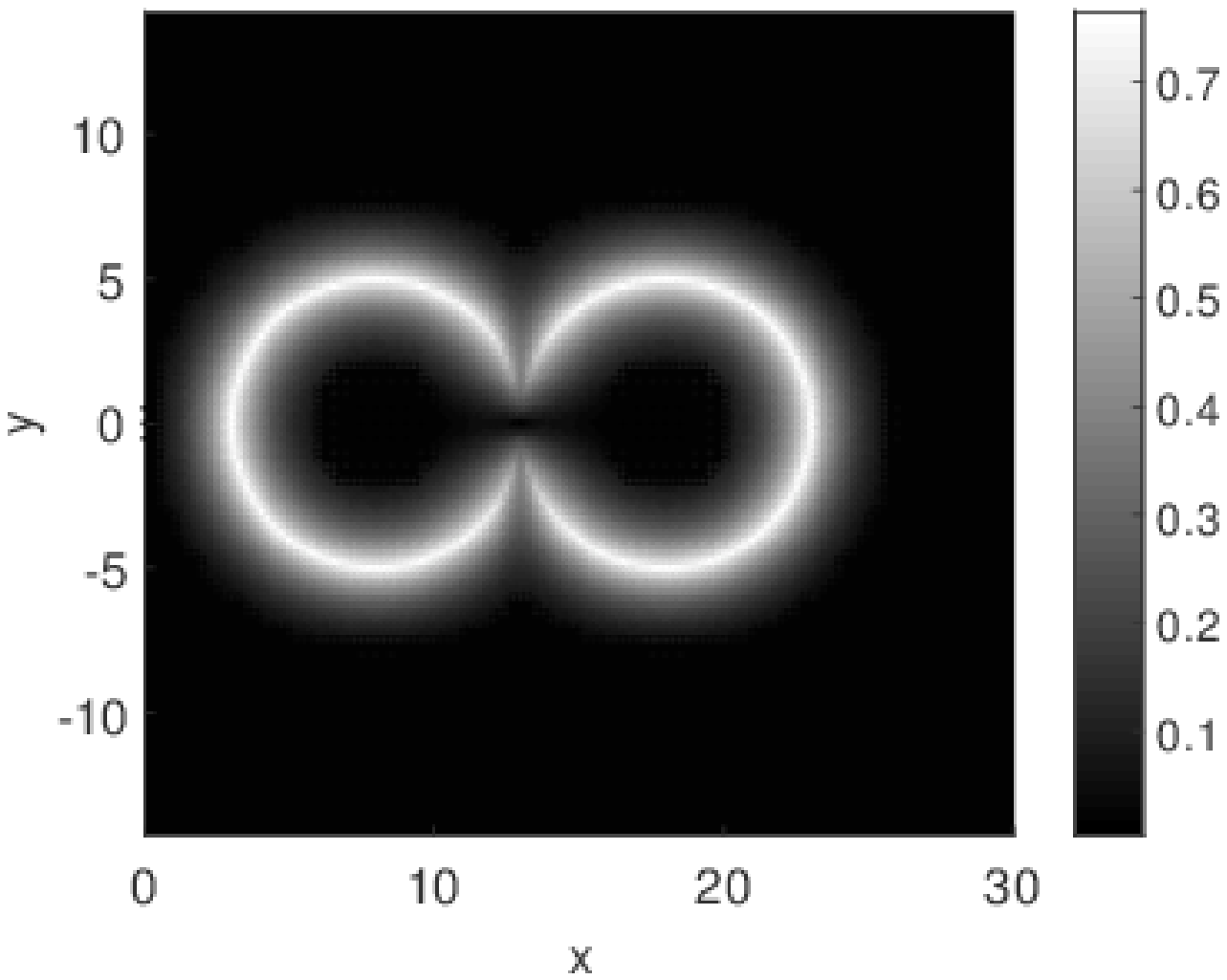}
\includegraphics[scale=0.5]{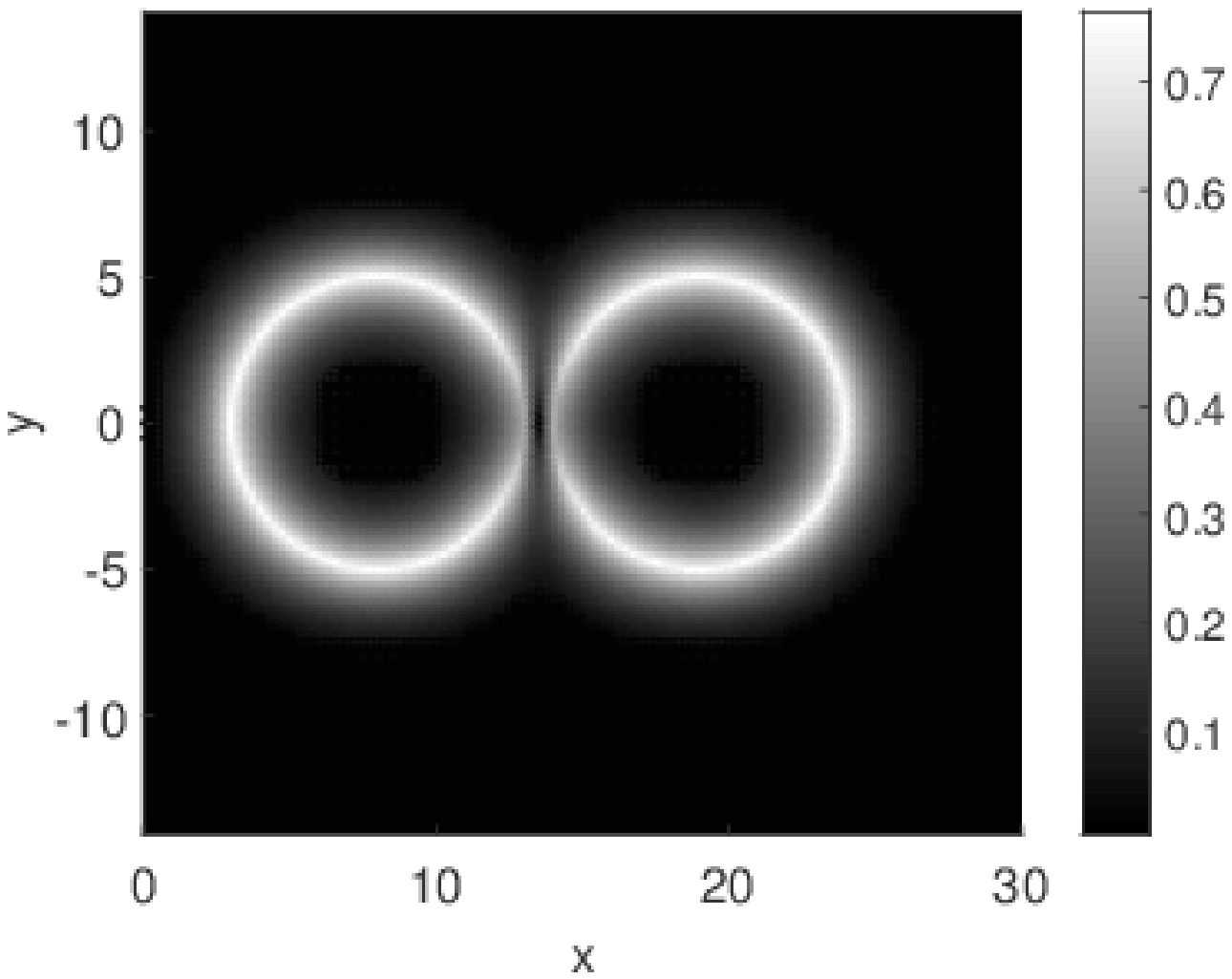}
\includegraphics[scale=0.5]{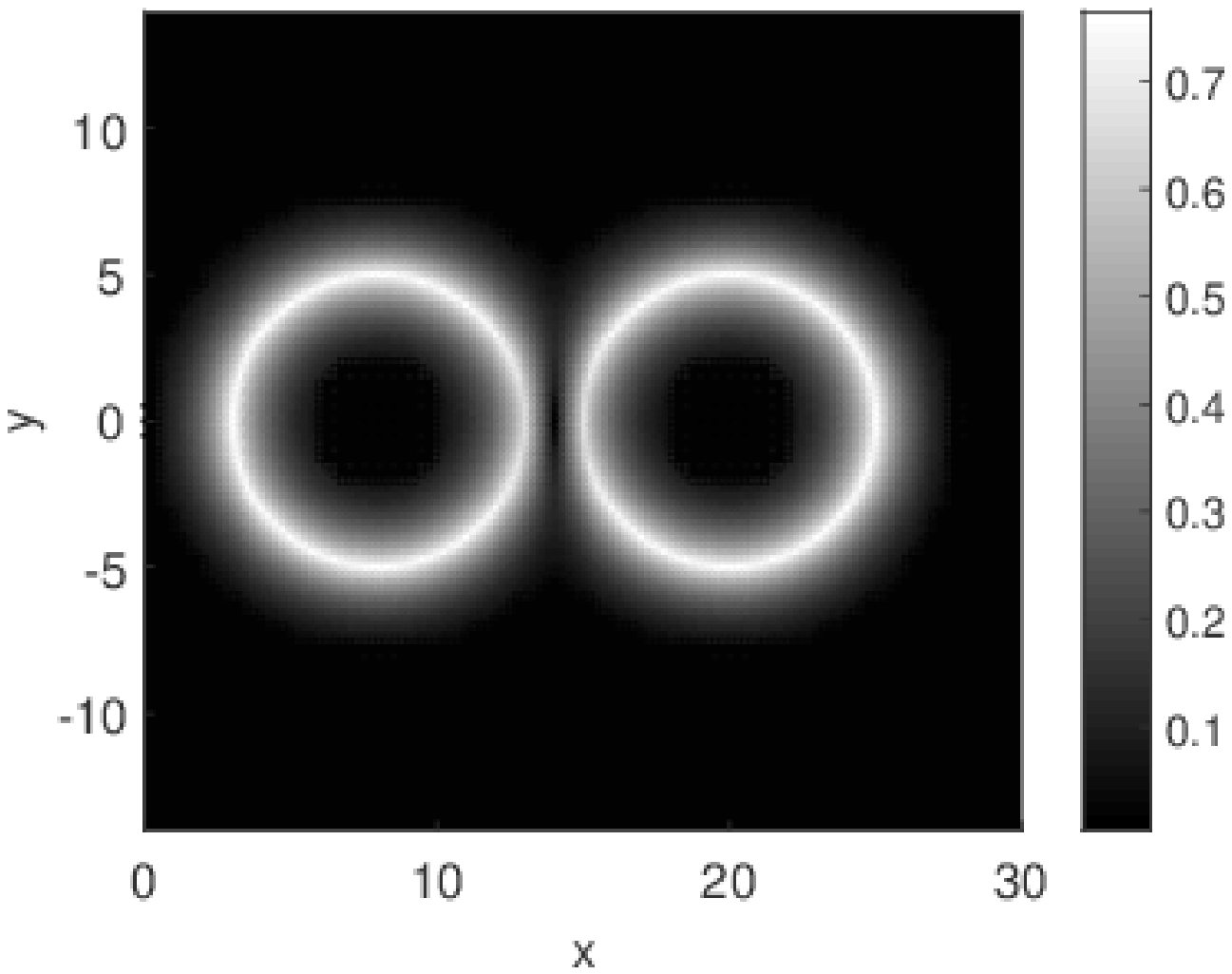}
\includegraphics[scale=0.5]{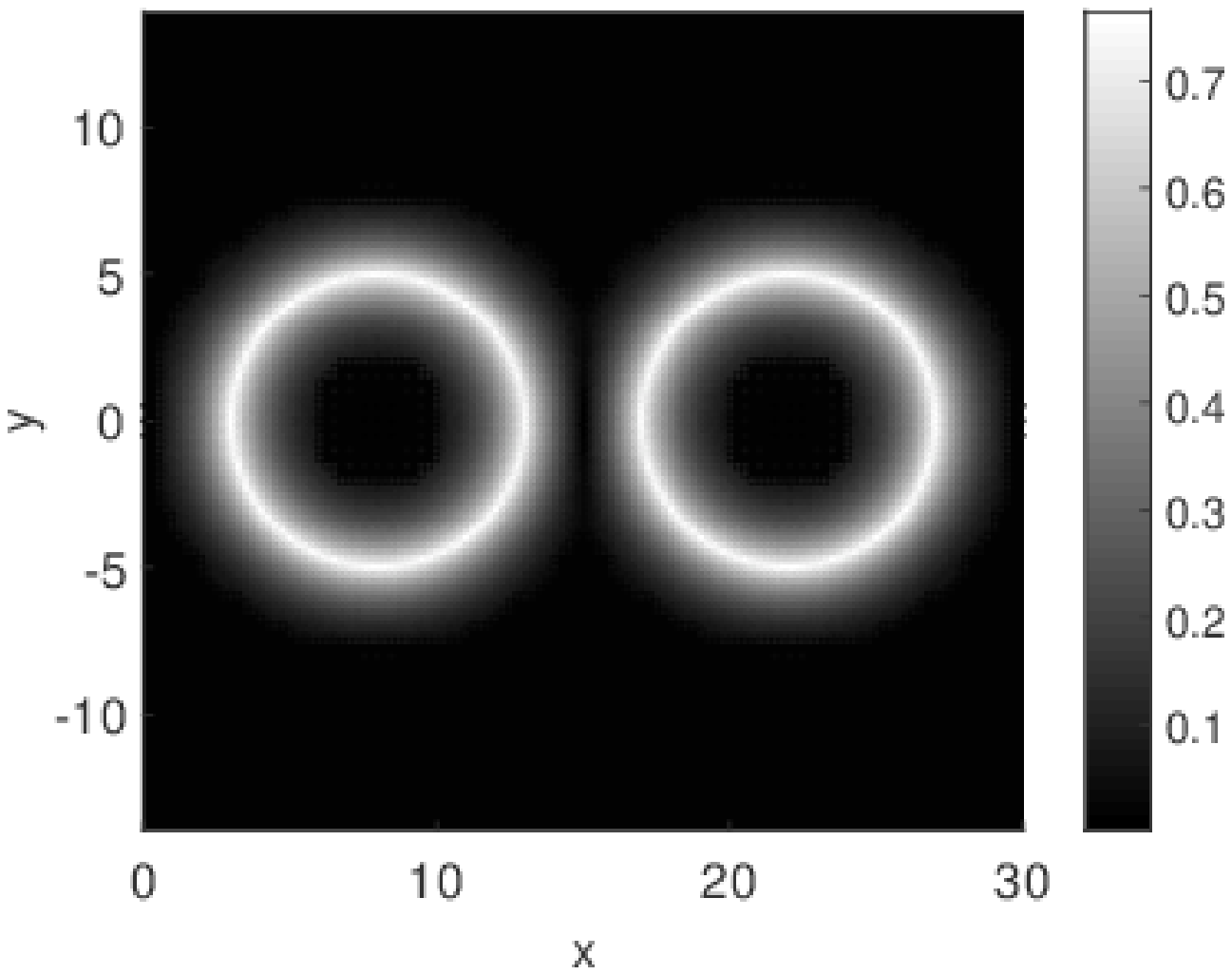}
\raggedleft{}\protect\caption{ Tilt angle $\theta$ profile for a system of two rafts with radius $R=5$ and
$\qb=-\qr=0.71$, for four values of the edge-to-edge separation $D$ of the rafts (which is given by their center-to-center separation minus the raft diameter). The value of $\theta$ is indicated by the grayscale bar to the right of each figure. Here we choose $\lamr=\lamb=\lambda$ and measure lengths in units of $\lambda$.  The figure shows a portion of the total system, which is a square of side $L=60$. The maximum tilt angle $\theta_{0}$ occurs at the edge of the rafts. (a) $D=0$, (b) $D=1$, (c) $D=2$,
(d) $D=4$. \label{fig:tilt-angle-}}
\end{figure}

We now consider the more complex and interesting situation of two rafts composed of molecules of the same chirality fixed in a background membrane of opposite (but not necessarily equal) chirality. We relax the constraints $\lamb=\lamr$ (i.e., $\Cb=\Cr$) and $|\qb|=|\qr|$ so that we can explore the parameter space more fully, and compare our results with data from the experiments \cite{key-41}.  As in the case of a single raft, we use adaptive mesh refinement near the edges of the rafts, and compute the tilt angle and raft interaction potential as functions of the edge-to-edge separation $D$ of the rafts measured in units of $\lamb$ (we explain below why we choose to measure lengths in units of $\lamb$ rather than $\lamr$). Figure \ref{fig:tilt-angle-} shows grayscale plots for the tilt angle computed by minimizing $f_\text{n}$ for two rafts of radius $R=4$ at several values of $D$  for the case  $\lamb=\lamr$. The lightest rings in the figure (corresponding to the maximum value of $\theta$) occur near the edges of the rafts. The figure illustrates the origin of the repulsive force between rafts already discussed in Ref.~\cite{key-41}. When the rafts are far apart (Figs.~\ref{fig:tilt-angle-}~c, d), the regions of maximum $\theta$ are full circles centered on the raft centers. As in the case of a single raft (see Fig.~\ref{fig:Plots-of-tilt}), starting from the center of one of the rafts and moving radially outward, the tilt angle grows until reaching the edge of the raft, then decreases as one moves into the background membrane where the viruses have opposite chirality. As the second raft is approached, the tilt angle begins to grow again until it reaches the edge of the raft, subsequently decreasing as one enters the raft interior. Thus, the preferred chiral twist in both the rafts and the background near their interfaces is satisfied. However, when the edges of the rafts are separated by a distance less than $2\lamb$ the preferred twist is no longer fully satisfied, as indicated in Figs.~\ref{fig:tilt-angle-}a, b.

\subsubsection{ Dependence of the interaction strength on control parameters} 
\begin{figure}
\includegraphics[scale=0.5]{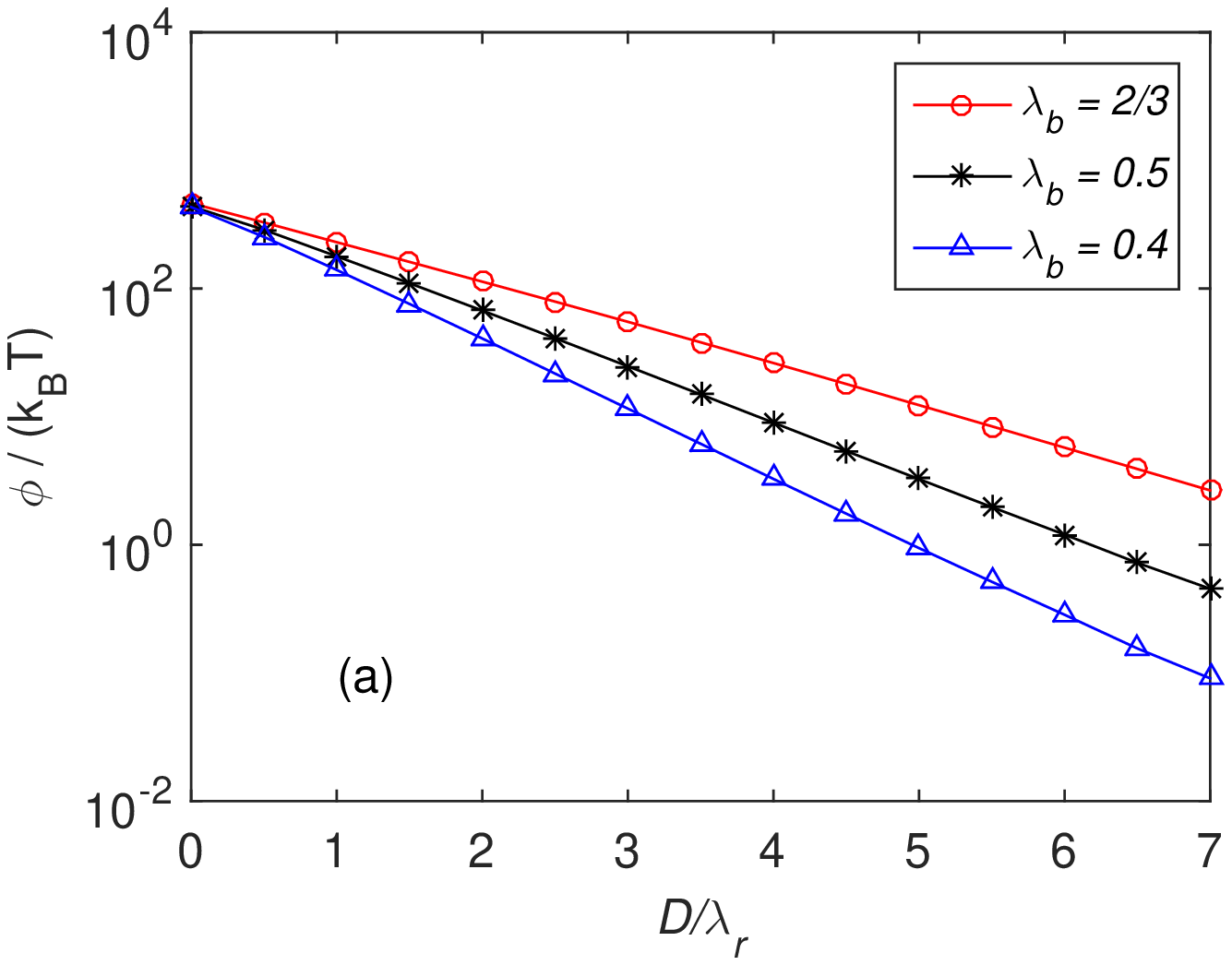}
\includegraphics[scale=0.5]{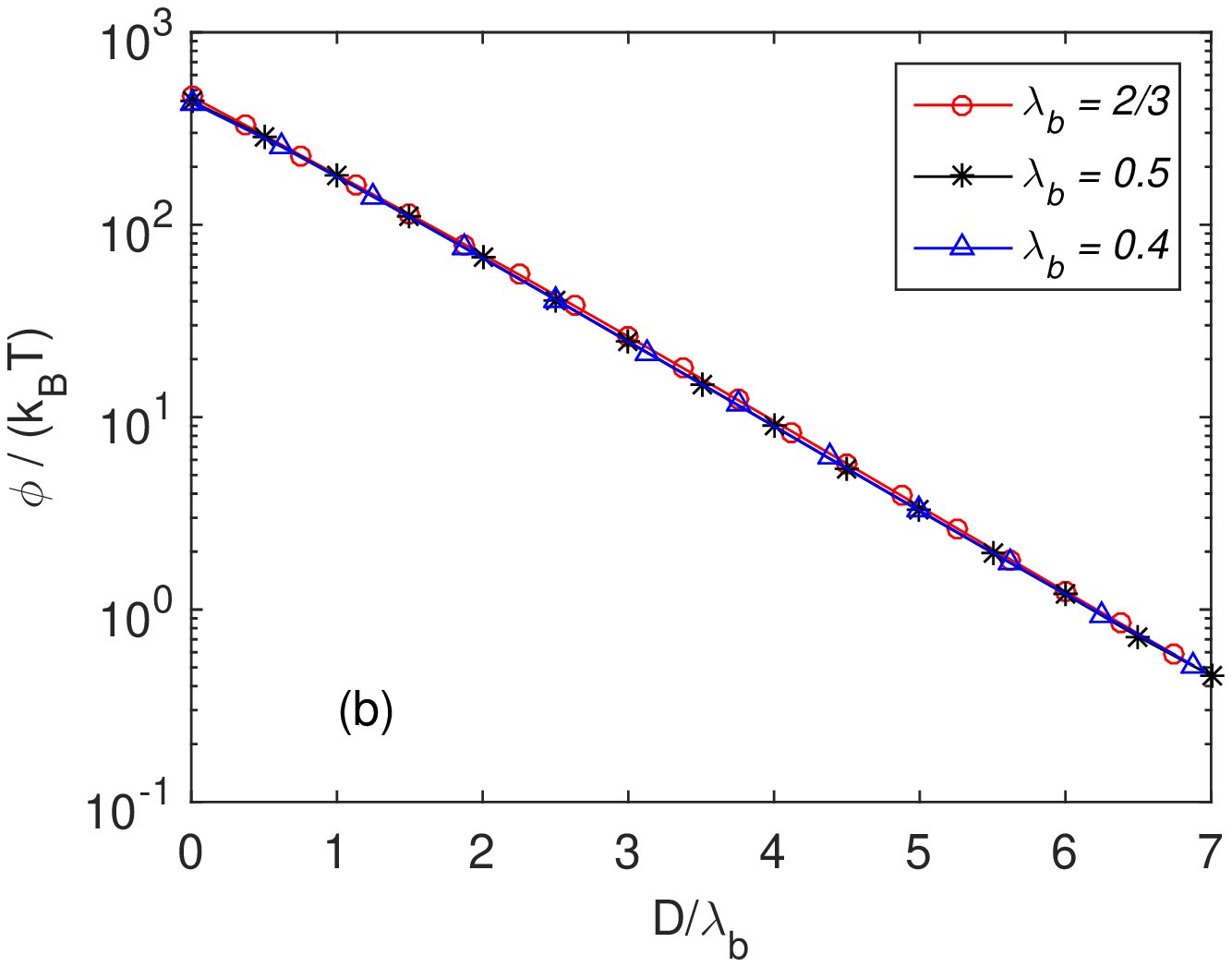}
\includegraphics[scale=0.5]{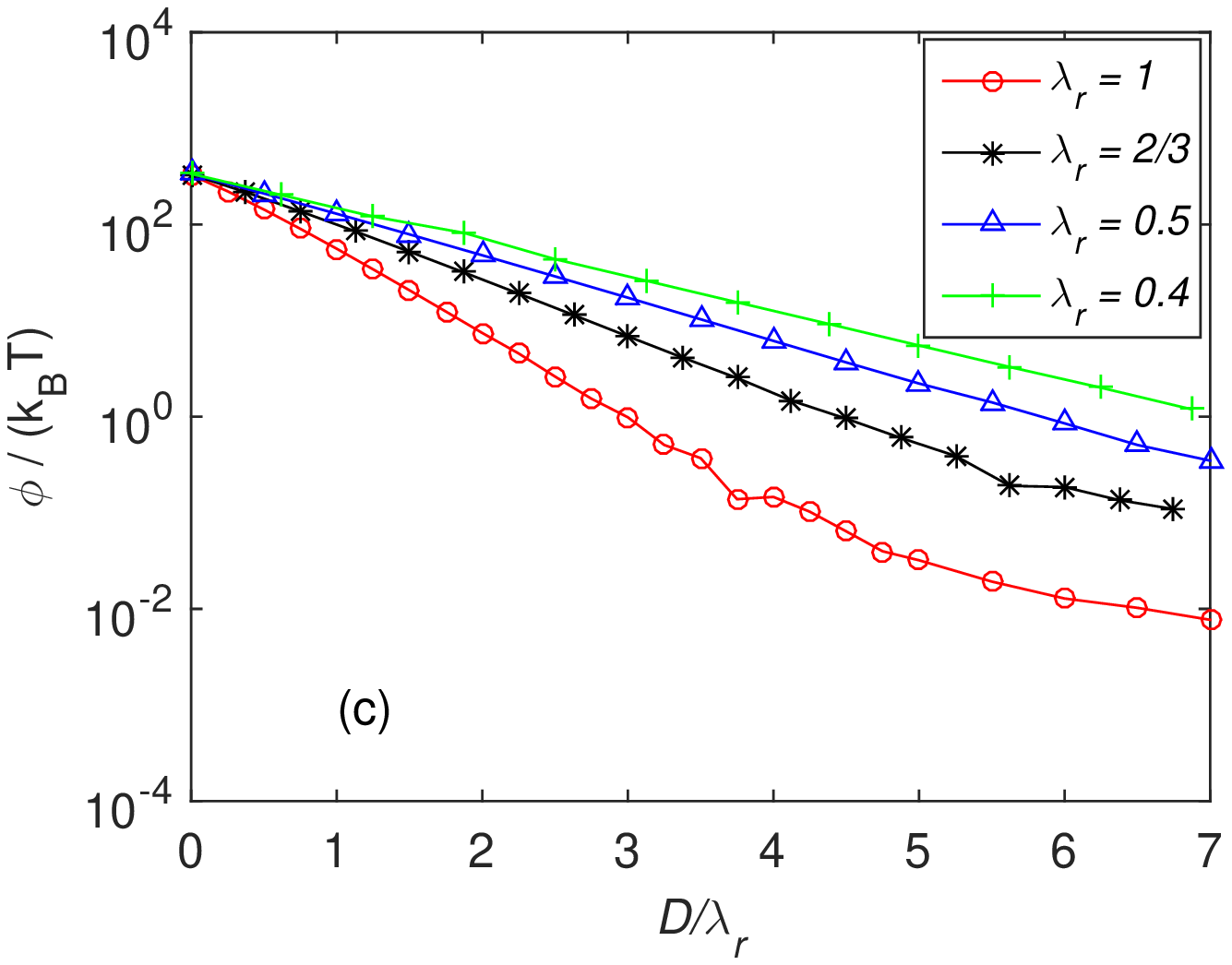}
\includegraphics[scale=0.5]{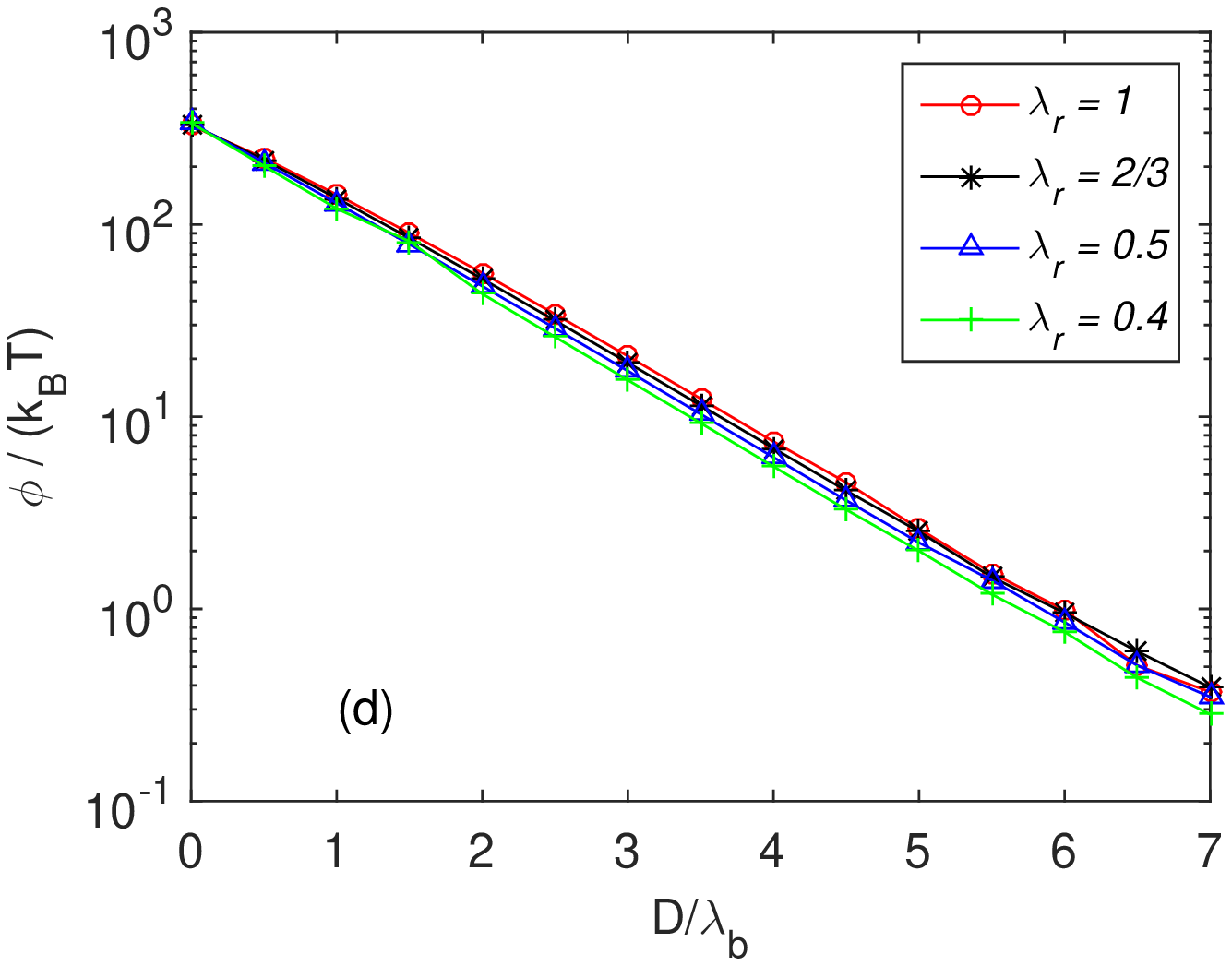}
\protect\caption{ Raft interaction potential $\phi$ (units of $k_B T$) as a function of edge-to-edge separation. (a) and
(b):  $\lamr=0.5$,  $R=3\lamr$; in (a) all lengths
are in units of $\lamr$, in (b) all lengths are in units of $\lamb$; (c) and (d): $\lamb=0.5$, $R=3\lamb$; in (c) all lengths are in units of $\lamr$; in (d) all lengths are in units of $\lamb$. Note the collapse of the curves in (b) and (d) where the unit of length is $\lamb$.   \label{fig:Rafts-proof-lambda_b}}
\end{figure}

The above discussion suggests that the characteristic length scale of the raft interaction is $\lamb$, the twist penetration depth in the background membrane. We confirm this assertion by comparing the interaction as a function of the raft separation in units of $\lamr$ and in units of $\lamb$ .  Figures \ref{fig:Rafts-proof-lambda_b}a and \ref{fig:Rafts-proof-lambda_b}c show the raft interaction measured in units of $\lamr$.  In Fig. \ref{fig:Rafts-proof-lambda_b}a the values of $\lamr$ and $R$ are held fixed while $\lamb$ is varied; in Fig.~\ref{fig:Rafts-proof-lambda_b}c the value of $\lamb$ is held fixed while $\lamr$ and $R$ are varied (such that $\lamr R$ is held fixed). Figures \ref{fig:Rafts-proof-lambda_b}b and \ref{fig:Rafts-proof-lambda_b}d show the raft interaction measured in units of $\lamb$ . In Fig. \ref{fig:Rafts-proof-lambda_b}b the value of $\lamr$ is held fixed while $\lamb$ and $R$ are varied (such that $\lamb R$ is held fixed); in Fig. \ref{fig:Rafts-proof-lambda_b}d  the values of $\lamb$ and $R$ are held fixed while $\lamr$ is varied. All other parameters are kept fixed. From Figs.~\ref{fig:Rafts-proof-lambda_b}b and \ref{fig:Rafts-proof-lambda_b}d we can see clearly that the interaction curves for rafts of different radii and values of $\lamr$ collapse when we measure lengths in units of $\lamb$.

Given that the raft interaction is mediated by the twist in the intervening background, we expect the range of the interaction to be determined by the distance over which the tilt angle in the background is appreciably larger than zero. As we saw in Fig. \ref{fig:Plots-of-tilt} this distance is approximately 5 (in units of $\lamb$), and we see from Figs. \ref{fig:Rafts-proof-lambda_b}b and \ref{fig:Rafts-proof-lambda_b}d that the interaction potential $\phi$ has dropped by an order of magnitude when $D\approx 5 \lamb$.

\begin{figure}
\includegraphics[scale=0.5]{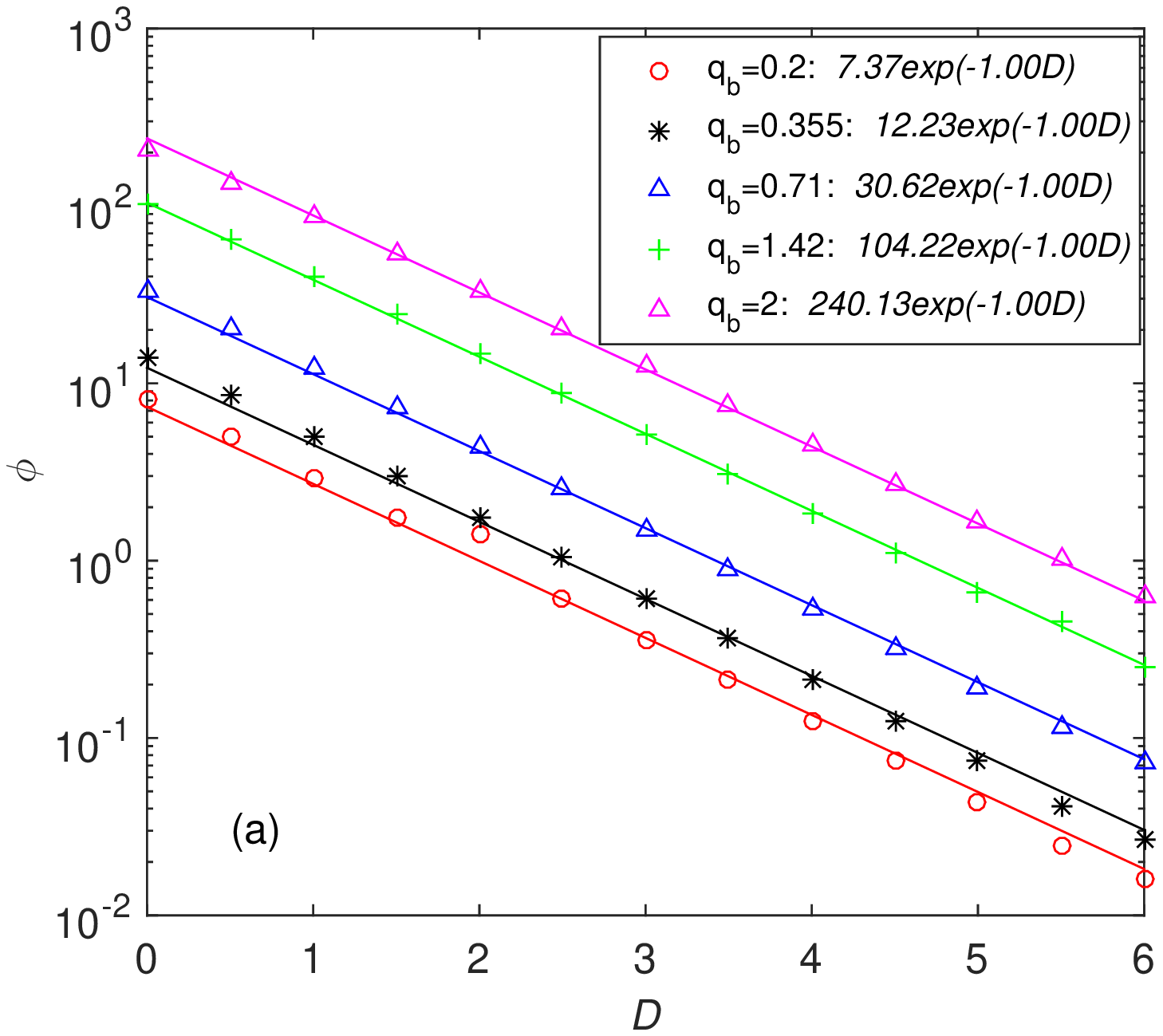}
\includegraphics[scale=0.5]{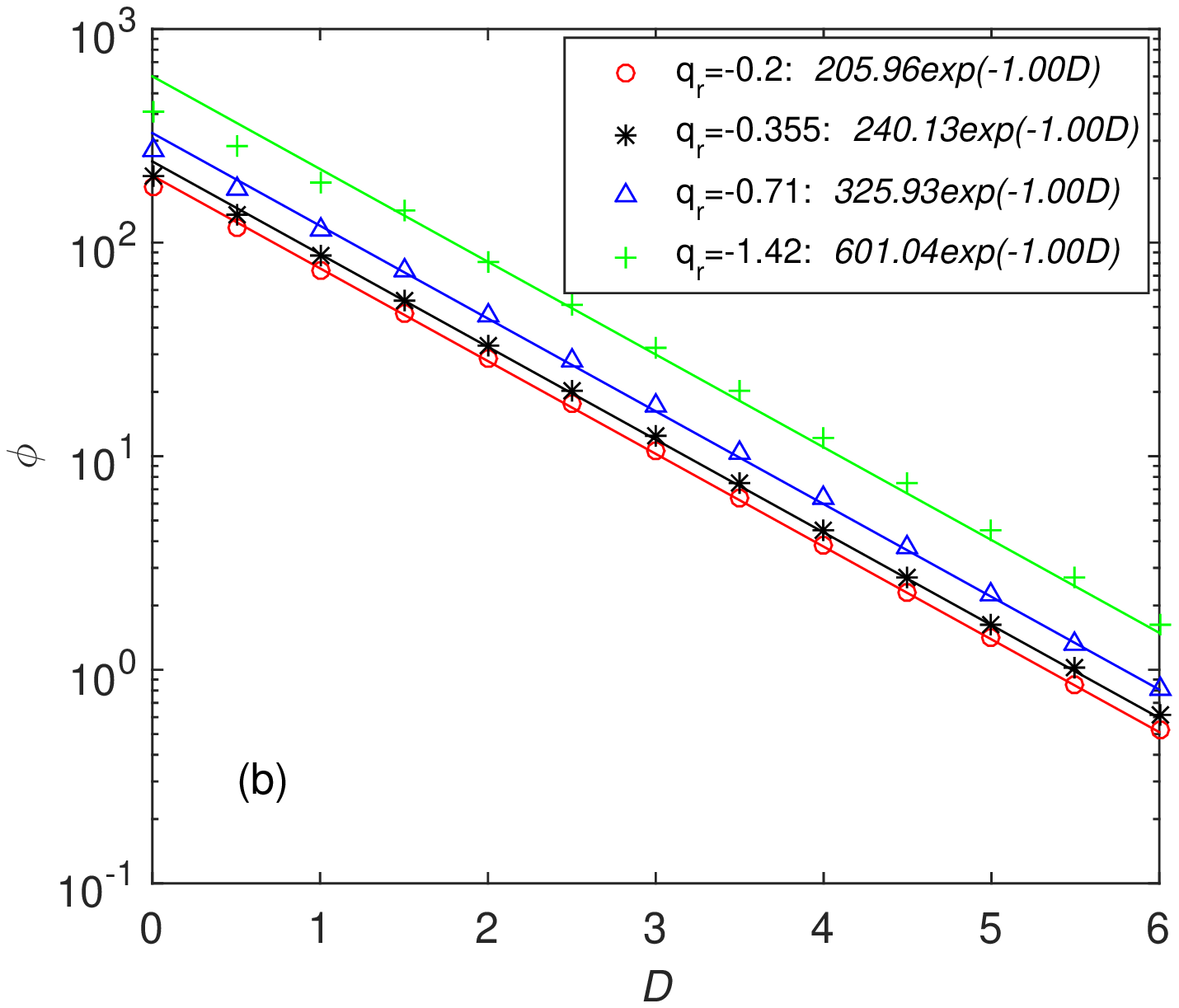}
\includegraphics[scale=0.5]{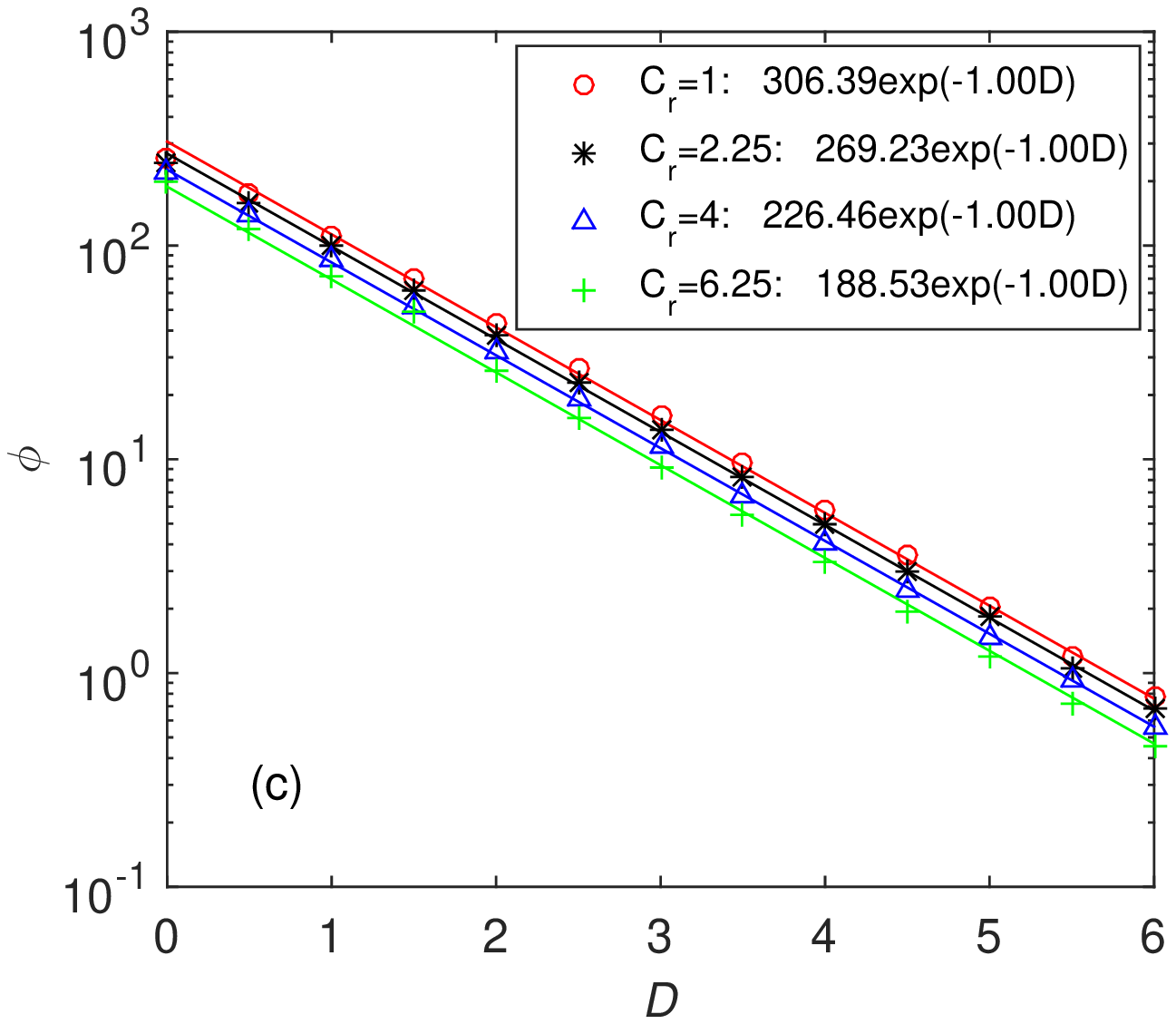}
\includegraphics[scale=0.5]{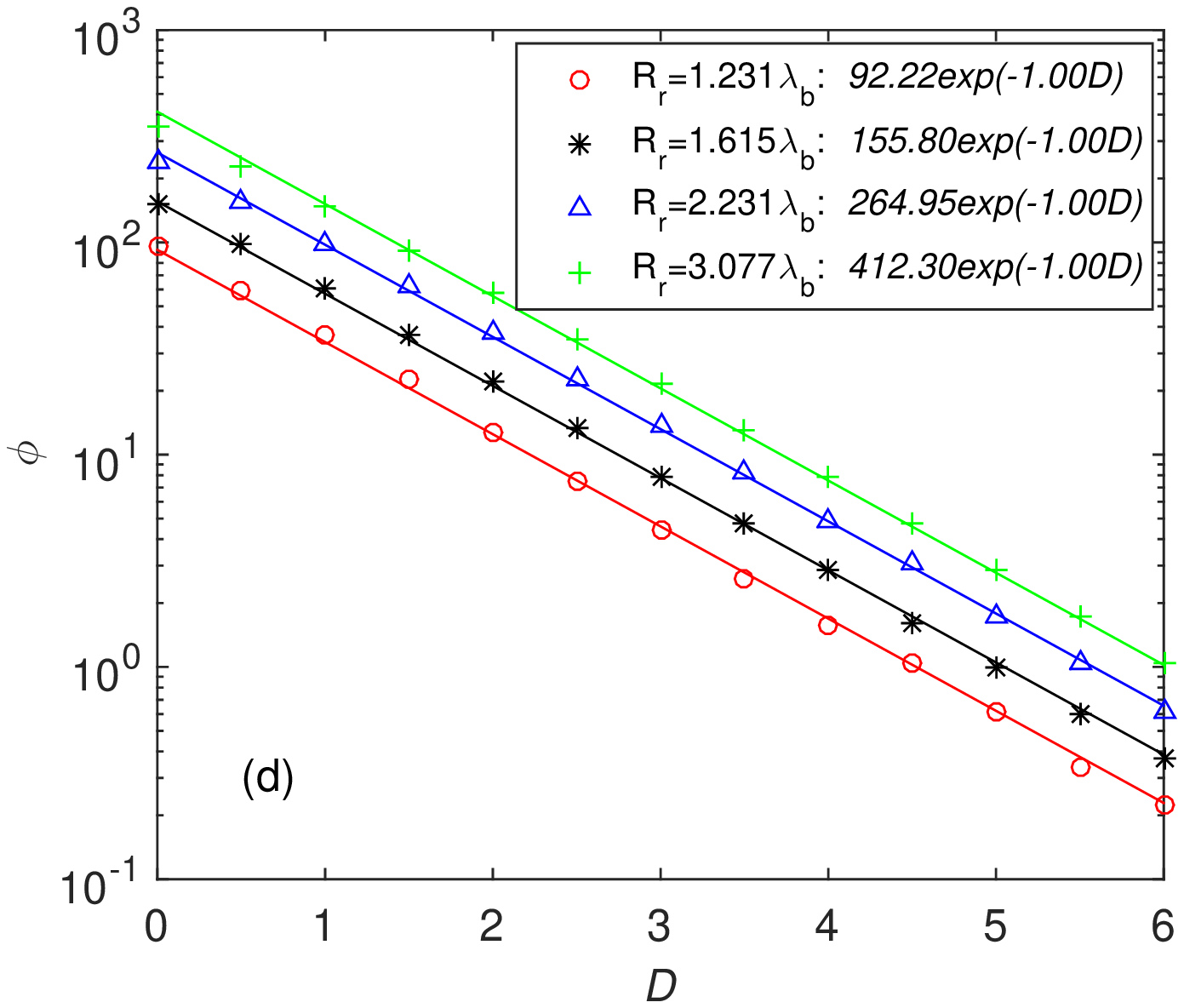}
\protect\caption{ Dependence of raft interaction $\phi$ (units of $k_B T$) on $\qb$, $\qr$, $\Cr$ and $R$. All lengths are measured in units of $\lamb$ and $\Cb=4$. The exponential expressions in the legend are fits to the data.  (a) $\qr=-0.355$, $\lamr=1.333$, $R=2.231$. b) $\qb=2$, $\lamr=1.333$, $R=2.231$. c) $\qr=-1.42$, $\qb=1.42$, $R=2.231$. d) $\qr=-1.42$, $\qb=1.42$, $\lamr=1.333$.   \label{fig:Rafts-interaction-strength}}
\end{figure}

Since the interaction range is determined by $\lamb$, we keep $\lamb=1$ throughout the following analysis and consider  the dependence of the interaction strength on $\qb$, $\qr$, $\Cr$ and $R$.  Figure \ref{fig:Rafts-interaction-strength} shows the effects of these
parameters on the raft interaction, by plotting on a semi-log scale the raft interaction as a function of $D$. In each case, the data is fit to an exponential form with decay length 1 (in units of $\lamb$).  We see that all of these parameters have very little effect on the decay length,  supporting the assertion made above that $\lamb$ determines the range of the interaction.
 From the figure we see that increasing the values of $\qb$, $|\qr|$ or $R$ increases the strength of the interaction, while increasing the value of  $\Cr$ decreases the strength. This result is readily understood in physical terms: increasing $\qb$ or $|\qr|$ leads to a larger value of the tilt angle at the raft edge, increasing  $R$ leads to a greater area of overlapped twisted domains for a given edge-to-edge distance, while increasing $\Cr$ leads to a smaller value of the tilt angle at the raft edge.

\begin{figure}
\includegraphics[scale=0.5]{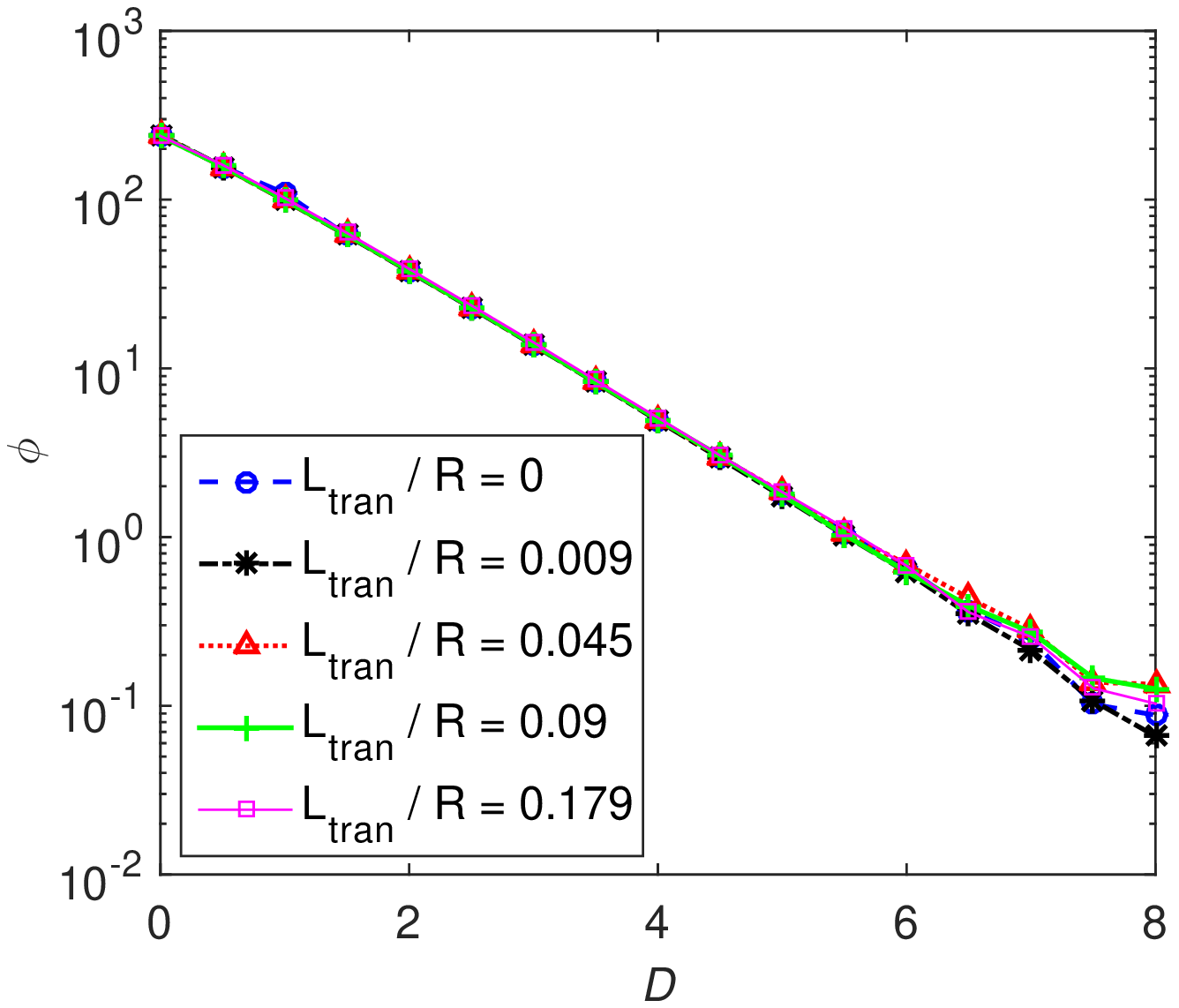}
\caption{Raft interaction potential $\phi$ (units of $\kt$) of two rafts of radius $R=2.231$ (units of $\lamb$) for five different linear chirality profiles where the chirality varies linearly from $\qr=1.42$ to $\qb=-1.42$ over a distance $2 \Lt$ centered on the edge of the raft. Here $\Cr=2.25$ and $\Cb=4.0$. \label{interaction-chirality}}.
\end{figure}

\begin{figure}
\includegraphics[scale=0.5]{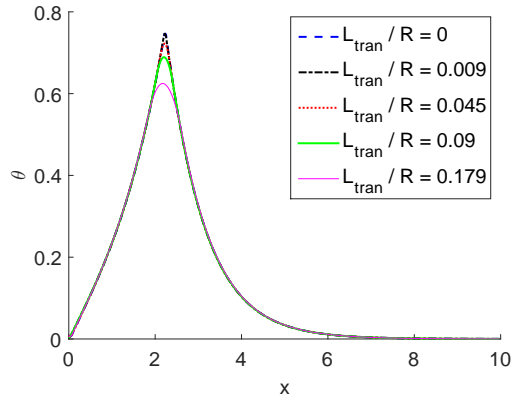}
\caption{ Tilt angle $\theta$ in a single raft of radius $R=2.231$ (in units of $\lamb$) as a function of the radial coordinate (units of $\lamb$) measured from the center of the raft for five different chirality profiles at the raft edge. The chirality profile is assumed to be linear with a width $2\Lt$ centered at $x=R$. In each case $\qr=1.42, \qb=-1.42, \Cr=2.25$ and $\Cb=4.0$.
\label{fig:tilt-chirality-profile}}
\end{figure}

In Sec. \ref{sub:Single-raft} we  assumed that the chirality jumped discontinuously  from $\qr$ to $\qb$  at the raft edge. In studying the interaction between two rafts, we have also considered a linear chirality profile of width $2\Lt$ centered on the edge of the raft, exploring five values of $\Lt/R$ (Fig.~\ref{interaction-chirality}). We find that the interaction is insensitive to the value of $\Lt/R$ except at large raft separations, $D\gtrsim 7$. In figs.~\ref{fig:Rafts-proof-lambda_b}, \ref{fig:Rafts-interaction-strength} and \ref {fig:Chirality-boundary-conditions-1} we have set $\Lt/R=0.045$.   We have also considered the effect of the chirality profile on the tilt angle. Figure \ref{fig:tilt-chirality-profile} shows that the tilt angle at the edge of the raft changes by more than 15\% as $\Lt$ is varied, a surprising result given the insensitivity of the raft interaction potential to the width of the chirality transition.

\subsubsection{ Comparison with experiments} 
\begin{figure}
\includegraphics[scale=0.5]{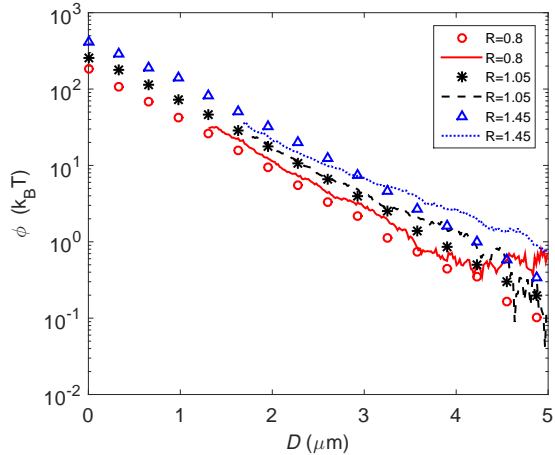}
\protect\caption{ Comparison of theoretical results for the raft interaction potential $\phi$ (symbols) with experimental data (lines) \cite{key-41}. Here, $\qr=-1.42$,
$\qb=2$, $\lamr=1.33\lamb$, chosen by eye to provide the best fit. Theoretical units were converted to physical units using $\lamb=0.65\mu m$, $K=125 k_BT$.  \label{fig:Chirality-boundary-conditions-1}}
\end{figure}

To compare our theoretical results with the experimental data of Ref.~\cite{key-41}, we convert our units to physical units using $\lamb=0.65\mu$m and $K=125\kt$ \citep{key-41}. The comparison is shown in Fig. \ref{fig:Chirality-boundary-conditions-1}, where we plot the potential $\phi$ as a function of $D$.
Since in the experiments the background rods show stronger chirality
and a shorter penetration depth than raft rods, we set $|\qb|>|\qr|$
and $\lamr>\lamb$, and chose values by eye that provided the best fit. The theoretical results show reasonable agreement with the data. While there is at present not sufficient data to stringently test the validity of a fit with three parameters, the ability of the theory to predict the functional form of the interaction profile and the trend of the interaction strength with varying $R$ is insensitive to the adjustable parameter values.

A discrepancy between theory and experiment is that rescaling the experimentally measured interaction potentials by the maximum tilt angle $\theta_{0}$ collapses the results onto a single curve \citep{key-41}, whereas rescaling the theoretical interaction potentials in this manner brings the curves closer together but does not result in collapse. However, we note that the theoretical prediction of the boundary tilt angle is sensitive to the imposed chirality profile (Fig.~\ref{fig:tilt-chirality-profile}). Moreover, since the experimental twist penetration length (over which most of the twist variation occurs) is on the order of the diffraction limit, it is not clear whether the theoretical value of $\theta_0$ can be directly mapped onto the experimental value. Thus, more information about the structure of the experimental raft edges may be required to address this discrepancy.

\section{Conclusions}
\label{conclusions}

Motivated by experiments \cite{key-41} on colloidal membranes exhibiting chiral rafts floating in a background membrane, we have theoretically studied the director-mediated interaction between two such rafts. We numerically minimized the Frank free energy in the single elastic constant approximation on a two-dimensional mesh using the Newton-Gauss-Seidel method with adaptive mesh refinement near the raft edges. We computed the interaction between the rafts by comparing the free energy of the system (rafts plus background membrane) with a reference system in which the rafts are far apart and noninteracting. In agreement with the interpretation suggested in  Ref.~\cite{key-41}, we found that the characteristic length scale of the interaction is the chiral penetration depth of the background membrane.  In particular, an isolated chiral raft in a background of rods with opposite chirality is stabilized by twisting of the background rods. The decay of this twist with distance from the raft is set by the membrane chiral penetration depth. Juxtaposing two rafts with an edge-to-edge separation less than twice the background penetration depth perturbs the preferred chiral twist in the background membrane between them, leading to a repulsive interaction.
As in the experiments, we found that the strength of the raft interaction increases with increasing raft diameter.   One discrepancy between our theory and the experiments is that Sharma et al. \cite{key-41} found that rescaling the interaction by the tilt angle at the edge of a raft collapses the results for different sized rafts on to a single curve. Our theoretical curves come closer together upon rescaling, but do not fully collapse.

\appendix

\section{\label{appendix} Discretizing the free energy}
We discretize $f_{n}$, Eq. (\ref{fn}),  on a square grid using the finite difference method for first and second order partial derivatives of $\theta$ and $\varphi$ with respect to $x$ and  $y$. We consider three neighboring points along a coordinate direction to evaluate derivatives at the central point. We use the symmetric derivative approximation to the first derivative and the central difference approximation to the second derivative. However, because of the periodicity of $\theta$  and $\varphi$ and the equivalence of  $\hat{\mathbf n}$ and $-\hat{\mathbf n}$, some care is required as we now discuss.

Let $\hat{\mathbf n}_{1}$, $\hat{\mathbf n}$ and $\hat{\mathbf n}_{2}$ be directors at
three neighboring grid points $A$, $B$ and $C$ located from left
to right on the $x$ axis and we wish to evaluate the derivatives at the central point $B$.  For a fixed value of the central director $\hat{\mathbf n}$,
four possible rotations may occur for the three points: (1) $\hat{\mathbf n}_{1}$
to $\hat{\mathbf n}$ to $\hat{\mathbf n}_{2}$, (2) $-\hat{\mathbf n}_{1}$ to $\hat{\mathbf n}$
to $\hat{\mathbf n}_{2}$, (3) $\hat{\mathbf n}_{1}$ to $\hat{\mathbf n}$ to $-\hat{\mathbf n}_{2}$,
(4) $-\hat{\mathbf n}_{1}$ to $\hat{\mathbf n}$ to $-\hat{\mathbf n}_{2}$. Among the four,
we select the one that minimizes the total rotation angle per unit distance and thus the Frank energy.
For example, for candidate (1), we compute the angle $\alpha$ between
$\hat{\mathbf n}_{1}$ and $\hat{\mathbf n}$, and the angle $\beta$ between $\hat{\mathbf n}$
and $\hat{\mathbf n}_{2}$. Then the total rotation angle per unit distance equals $\alpha/\overline{AB}+\beta/\overline{BC}$,
where $\overline{AB}$, $\overline{BC}$ are the distances between the neighboring grid points  (these distances may not be equal due to mesh refinement).
The director  $-\hat{\mathbf n}$ corresponds to  $\theta\rightarrow \pi-\theta, \varphi \rightarrow\varphi+\pi$. Then we get the
derivatives for $\theta$ directly by finite difference method. For
$\varphi$, we need to consider $\varphi+2\pi$ and $\varphi-2\pi$
as well, and we select the candidate which minimizes the sum of the
first and second order derivatives.

\section{Tilt angle at interface between two semi-infinite membranes of different chiraility}

In this appendix we calculate exactly the value of the director tilt angle at the interface between two semi-infinite membranes of different chirality. If the membranes have equal and opposite chirality we find that the tilt angle has the same value as when a single semi-infinite membrane is bounded by a polymer depletant and zero director torque is assumed at the interface \cite{Barry2009}.

We consider two semi-infinite membranes lying in the $x-y$ plane and in contact along the $y$ axis. The membrane in the $x>0$ half-plane has chirality $q_1>0$, while the one in the $x<0$ half-plane has chirality $q_2<0$. The symmetry of the semi-infinite geometry dictates that the director tilts by an angle $\theta$ in the $y$ direction only and the free energy density $f_\text{n}$ (Eq. (\ref{fn})) simplifies to:
\begin{equation}
f_\text{n}=\frac{K}{2}\biggl(\frac{d\theta}{dx}-q(x) \biggr)^2+\frac{C}{2}\sin^{2}\theta\label{fnB},
\end{equation}
where the local chirality $q(x)$ is specified above.
The corresponding Euler-Lagrange equation is given by:
\begin{equation}
\lambda^2\frac{d^2\theta}{dx^2} -\lambda^2\frac{dq}{dx}- \sin2\theta=0,
\label{EL}
\end{equation}
with $\lambda=\sqrt{K/C}$.  For the assumed discontinuous jump in the chirality we have:
\begin{equation}
\frac{dq}{dx} = (q_1 + |q_2|)\delta(x).
\end{equation}
Integrating Eq.~(\ref{EL}) from $x=-\epsilon$ to $x=\epsilon$ we find in the limit $\epsilon\rightarrow0$:
\begin{equation}
\frac{d\theta}{dx}\biggr\vert^{\epsilon}_{-\epsilon}=q_1+|q_2|.
\label{eq1}
\end{equation}
The first integral of the Euler-Lagrange equation is given by:
\begin{equation}
\lambda\frac{d\theta}{dx}=\pm \sin\theta,
\label{eq2}
\end{equation}
where the upper and lower signs correspond to $x>0$ and $x<0$, respectively.
Combining Eqs.(\ref{eq1}) and (\ref{eq2}) we obtain in the limit $\epsilon\rightarrow0$:
\begin{equation}
\sin\theta(0)=\frac{\lambda}{2} (q_1 + |q_2|),
\label{eq3}
\end{equation}
where we have assumed that $\theta$ is continuous at the interface, $x=0$.
From Eq. (\ref{eq3}) we see that if $q_1=-q_2=q$ then $\sin \theta(0) = \lambda q$ which is identical to the result found in Ref.~\cite{Barry2009} for a semi-infinite membrane bounded by a polymer depletant imposing zero director torque at the interface.

\begin{acknowledgments}

We thank Z. Dogic, R. Sakhardande and P. Sharma for helpful discussions.  This work was supported by the NSF through MRSEC Grant No. 1420382.

\end{acknowledgments}

\end{document}